# Off-diagonal magnetoimpedance in field-annealed Co-based amorphous ribbons

N. A. Buznikov,[a)] CheolGi Kim,[b)] Chong-Oh Kim, and Lan Jin

*Research Center for Advanced Magnetic Materials, Chungnam National University,*

*Daejeon 305-764, Korea*

Seok-Soo Yoon

*Department of Physics, Andong National University, Andong 760-749, Korea*

**Abstract**

The off-diagonal magnetoimpedance in field-annealed CoFeSiB amorphous ribbons was measured in the low-frequency range using a pickup coil wound around the sample. The asymmetric two-peak behavior of the field dependence of the off-diagonal impedance was observed. The asymmetry is attributed to the formation of a hard magnetic crystalline phase at the ribbon surface. The experimental results are interpreted in terms of the surface impedance tensor. It is assumed that the ribbon consists of an inner amorphous region and surface crystalline layers. The coupling between the crystalline and amorphous phases is described through an effective bias field. A qualitative agreement between the calculated dependences and experimental data is demonstrated. The results obtained may be useful for development of weak magnetic-field sensors.

[a)] Permanent address: Institute for Theoretical and Applied Electrodynamics, Russian Academy of Sciences, Moscow, Russia
[b)] Author to whom correspondence should be addressed; electronic mail: cgkim@cnu.ac.kr

# I. INTRODUCTION

The giant magnetoimpedance (GMI) effect consists in a huge change in the impedance of a soft magnetic conductor in the presence of a static magnetic field. The interest in the GMI is supported by possible use of the effect in various technological applications, in particular for the development of weak magnetic-field sensors. The linearity and the sensitivity for the magnetic field are the most important parameters in practical applications of the GMI effect. To improve the linear characteristic of the GMI response, the asymmetric GMI is very promising.[1,2]

There are three mechanisms of the asymmetric GMI.[2] The first one has been observed in Co-based amorphous wires with dc bias current superimposed on the driving current.[3,4] This type of asymmetry is caused by the combination of helical magnetic anisotropy with circular dc field produced by the bias current and is related to the asymmetry in the dc magnetization. Another method of producing the asymmetric GMI profile consists of applying an axial ac bias field to a sample.[5] In this case, the asymmetry results from the mixing of the diagonal and off-diagonal components of the impedance tensor due to the ac cross-magnetization process.[1] The third type of the asymmetric GMI has been observed in Co-based amorphous ribbons annealed in air in the presence of a weak magnetic field.[6–9] Since the ribbons annealed in vacuum did not show the asymmetry,[9] the phenomenon has been attributed to oxidation and surface crystallization. This type of annealing produces asymmetric hysteresis loops in amorphous ribbons due to the exchange interaction of amorphous bulk with the magnetically harder crystalline surface layers.[10] In the presence of the annealing field, unidirectional magnetic anisotropy is induced in the crystalline layers. Because the crystalline phase is the hard magnetic one, it remains magnetically ordered within relatively wide range of magnetic fields. The characteristics of the amorphous and crystalline phases in field-annealed Co-based amorphous ribbons have been analyzed separately by means of the magneto-optical Kerr effect, and the shift of loop for amorphous phase within low-field region has been revealed due to the exchange coupling.[11] The coupling between the crystalline and amorphous phases produces an effective bias field resulting in the asymmetry in the dc magnetization that is responsible for the asymmetric GMI in field-annealed ribbons.[6,9] At sufficiently low frequencies, the GMI profile exhibits a drastic steplike change in the impedance near zero field (the so-called "GMI valve"), and at high frequencies, the field dependence of the impedance shows asymmetric two-peak behavior.[9] It should be noted that the peaks at the GMI field dependence are reversed if the applied field exceeds some value of the order of several hundreds of Oe.[9]

Usually, the GMI is measured as the field-dependent voltage across a soft magnetic sample. Another method to detect the field-dependent signal consists in the use of the pickup coil wound around a sample. This method is based on the cross-magnetization process, which appears since the current induces an ac axial magnetization variation and, hence, the pickup coil voltage. The effect has been referred to as off-diagonal magnetoimpedance. Off-diagonal magnetoimpedance has been studied in detail for Co-based amorphous wires[12–17] and multilayered films.[18,19] It has been demonstrated that the off-diagonal magnetoimpedance may be preferable for sensor applications since it generates a linear voltage response with enhanced field sensitivity. However, up to now the effect has not been investigated in amorphous ribbons.

In this paper, we present a study of off-diagonal magnetoimpedance in field-annealed Co-based amorphous ribbons within the frequency range 0.1–1 MHz. The asymmetric two-peak behavior of the field dependence of the off-diagonal impedance was observed, which is related to the coupling between crystalline and amorphous phases. The experimental results are interpreted in terms of the field and frequency dependent surface impedance tensor.[1,13,16,20,21] It is assumed that the ribbon consists of an inner amorphous core and two outer crystalline layers. Neglecting a domain structure, the off-diagonal magnetoimpedance is calculated. A satisfactory agreement between the calculated field and frequency dependences of the off-diagonal impedance and experimental data is demonstrated.

## II. EXPERIMENT

Amorphous ribbons $Co_{66}Fe_4Si_{15}B_{15}$ prepared by the rapid solidification technique were annealed in air at a temperature of 380 °C during 8 h. A field of 3 Oe was applied along the longitudinal direction of the sample during the annealing. The width of the samples was 2 mm, and their thickness was 20 μm. For experiments, the ribbons of length 3 cm were used.

The schematic diagram of the experimental setup is shown in Fig. 1. The amplitude $I_0$ of the ac current was controlled to 1 mA by a resistor of 100 Ω connected with the sample in series, and the current frequency $f$ was varied from 0.1 to 1 MHz by a function generator. The longitudinal dc magnetic field $H_e$ created by a Helmholtz coil was changed from −40 to 40 Oe. Note that the direction of the positive field coincides with that of the annealing field. The pickup coil was wound around the ribbon. The pickup coil had 25 turns and was 5.5 mm in length. In the experiments, the amplitude $V_c$ of the voltage induced in the pickup coil was measured by a lock-in amplifier as a function of the frequency and longitudinal dc magnetic field. From the measured voltage amplitude, the off-diagonal impedance $Z_c$ was calculated as $Z_c = V_c/I_0$.

### III. RESULTS AND DISCUSSION

Figure 2 presents the measured field dependence of the off-diagonal impedance at different frequencies. The dependence exhibits asymmetric two-peak behavior, with the peak at a positive field being higher than the peak at a negative field. Similar to the case of the GMI response measured as the voltage across the field-annealed ribbons,[6–9] the asymmetry in the field dependence of the off-diagonal impedance may be attributed to the coupling between the amorphous phase and the surface crystalline layers appearing after annealing.

It follows from Fig. 2 that there is a hysteresis in the field dependence of the off-diagonal impedance for increasing and decreasing fields. The hysteresis was observed within the region of low external fields and may be related to the effect of the domain structure. It is seen from Fig. 2 that the off-diagonal magnetoimpedance increases with the frequency, and, in the same time, the asymmetry between peaks decreases. The field dependence of the off-diagonal impedance at low magnetic fields demonstrates almost linear behavior, which is promising for sensor applications. At $H_e=0$, the field sensitivity $\Delta Z_c/\Delta H_e$ attains 0.02 Ω/Oe, 0.05 Ω/Oe, and 0.14 Ω/Oe for 100 kHz, 500 kHz, and 1 MHz, respectively.

To interpret the observed dependences of the off-diagonal impedance, we use a simple model that allows one to describe the main features of the experimental results. The annealing leads to the crystallization of the surface layers of the ribbon.[7] Due to the presence of the annealing field, the crystallites are magnetically ordered, which results in an effective unidirectional surface anisotropy. The value of the unidirectional anisotropy field $H_u$ is sufficiently high and attains several hundreds of Oe.[9] In this analysis, we consider the ribbon of thickness $t$ consisting of an inner amorphous region and two outer crystalline layers at the ribbon surface.[22] It is assumed also that the surface layers have different thickness, $t_1$ and $t_2$. The difference in the thickness of the crystalline layers may be related to peculiarities of the field-annealing process. During the annealing, one side of the ribbon is located on the substrate, which may result in a change of the crystalline layer thickness in comparison with that at another side of the ribbon. It is assumed further that the unidirectional anisotropy field in the outer layers has the constant angle $\varphi$ with respect to the transverse direction (a sketch of the coordinate system is shown in Fig. 3). Note that the direction of $H_u$ differs from that of the annealing field, which may be attributed to the influence of the amorphous core on the crystallization process in the surface layers.[22,23]

The distribution of the magnetization in the amorphous core depends on the stresses induced in the ribbon during the fabrication process and may vary significantly over the ribbon volume. We assume for simplicity that the amorphous core has a single-domain structure, the uniaxial anisotropy field $H_a$ is constant, and the anisotropy axis makes the angle $\psi$ with the transverse direction in the whole amorphous region. The coupling between amorphous and crystalline phases induces the effective bias field $H_b$ in the amorphous region, with the bias field being in the opposite direction to the unidirectional anisotropy field $H_u$ (Ref. 9) (the angle of the bias field with respect to the transverse direction is $\varphi+\pi$, see Fig. 3). We assume for simplicity that the effective bias field does not vary over the amorphous region.

The analysis is concerned with the calculation of the surface impedance tensor,[1,13,16,20,21] which relates the components of the electric and magnetic field at the sample surface. The surface impedance calculation is based on a solution of Maxwell equations for the electric and magnetic fields together with the Landau–Lifshitz equation. The permeability tensor for both the inner amorphous region and outer crystalline layers can be found by means of the well-known procedure of the solution of the linearized Landau–Lifshitz equation, and the permeability is represented by a nondiagonal tensor.[1,13,15,16]

Since the ribbon length and width are much higher than its thickness, we may assume that the fields depend only on the coordinate perpendicular to the ribbon plane ($x$-coordinate). In this case, Maxwell equations in the amorphous region, $-t/2+t_2<x<t/2-t_1$, reduce to coupled differential equations for the transverse, $h_y^{(a)}$, and longitudinal, $h_z^{(a)}$, components of the ac magnetic field:[14,18]

$$\frac{d^2 h_y^{(a)}}{dx^2} + \frac{2i}{\delta_0^2}(1+\mu_1 \sin^2\theta) h_y^{(a)} = \frac{2i}{\delta_0^2} \mu_1 h_z^{(a)} \sin\theta \cos\theta,$$

$$\frac{d^2 h_z^{(a)}}{dx^2} + \frac{2i}{\delta_0^2}(1+\mu_1 \cos^2\theta) h_z^{(a)} = \frac{2i}{\delta_0^2} \mu_1 h_y^{(a)} \sin\theta \cos\theta,$$

(1)

where $\delta_0 = c/(2\pi\sigma\omega)^{1/2}$ is the skin depth in nonmagnetic material, $c$ is the velocity of light, $\sigma$ is the ribbon conductivity, $\omega = 2\pi f$ is the angular current frequency, and $\mu_1$ is the effective permeability of the amorphous region, which is given by

$$\mu_1 = \frac{\omega_m(\omega_m + \omega_1 - i\kappa\omega)}{(\omega_m + \omega_1 - i\kappa\omega)(\omega_2 - i\kappa\omega) - \omega^2},$$

$$\omega_m = \gamma 4\pi M,$$

$$\omega_1 = \gamma[H_a \cos^2(\theta - \psi) - H_b \cos(\theta - \varphi) + H_e \sin\theta],$$

$$\omega_2 = \gamma[H_a \cos\{2(\theta - \psi)\} - H_b \cos(\theta - \varphi) + H_e \sin\theta].$$

(2)

Here $M$ is the saturation magnetization, $\gamma$ is the gyromagnetic constant, $\kappa$ is the Gilbert damping parameter, and $\theta$ is the equilibrium angle between the magnetization vector and the transverse direction. The angle $\theta$ can be found by minimizing the free energy, which can be presented as a sum of the uniaxial anisotropy energy, the bias field energy, and the Zeeman energy. The minimization procedure results in the following equation for the equilibrium angle:[22]

$$H_a \sin(\theta - \psi)\cos(\theta - \psi) - H_b \sin(\theta - \varphi) - H_e \cos\theta = 0.$$

(3)

Since the unidirectional anisotropy field $H_u$ in the crystalline layers is high, we may assume that the direction of the magnetization in the surface layers is almost independent of the external field at low $H_e$ and coincides with the direction of $H_u$.[22] In this case, equations for the ac magnetic field in the surface layers can be written in the form

$$\frac{d^2 h_y^{(j)}}{dx^2} + \frac{2i}{\delta_0^2}(1 + \mu_2 \sin^2 \varphi) h_y^{(j)} = \frac{2i}{\delta_0^2} \mu_2 h_z^{(j)} \sin \varphi \cos \varphi,$$
$$\frac{d^2 h_z^{(j)}}{dx^2} + \frac{2i}{\delta_0^2}(1 + \mu_2 \cos^2 \varphi) h_z^{(j)} = \frac{2i}{\delta_0^2} \mu_2 h_y^{(j)} \sin \varphi \cos \varphi,$$
(4)

where the superscripts $j=1,2$ correspond to the regions $t/2-t_1 < x < t/2$ and $-t/2 < x < -t/2+t_2$, respectively, and the effective permeability $\mu_2$ of the hard magnetic layers can be presented in the form

$$\mu_2 = \frac{\omega_m(\omega_m + \omega_3 - i\kappa\omega)}{(\omega_m + \omega_3 - i\kappa\omega)(\omega_3 - i\kappa\omega) - \omega^2},$$
$$\omega_3 = \gamma H_u.$$
(5)

The components of the magnetic field at the ribbon surface are determined by the excitation conditions and can be expressed as

$$h_y^{(1)}\Big|_{x=t/2} = 2\pi I(t)/cw, \quad h_z^{(1)}\Big|_{x=t/2} = 0,$$
$$h_y^{(2)}\Big|_{x=-t/2} = -2\pi I(t)/cw, \quad h_z^{(2)}\Big|_{x=-t/2} = 0,$$
(6)

where $I(t)$ is the current and $w$ is the ribbon width. Furthermore, the magnetic field should satisfy the continuity conditions at the interfaces between the amorphous region and crystalline layers:

$$h_y^{(a)}\Big|_{x=t/2-t_1} = h_y^{(1)}\Big|_{x=t/2-t_1}, \quad h_z^{(a)}\Big|_{x=t/2-t_1} = h_z^{(1)}\Big|_{x=t/2-t_1},$$
$$h_y^{(a)}\Big|_{x=-t/2+t_2} = h_y^{(2)}\Big|_{x=-t/2+t_2}, \quad h_z^{(a)}\Big|_{x=-t/2+t_2} = h_z^{(2)}\Big|_{x=-t/2+t_2},$$
$$\frac{dh_y^{(a)}}{dx}\Big|_{x=t/2-t_1} = \frac{dh_y^{(1)}}{dx}\Big|_{x=t/2-t_1}, \quad \frac{dh_z^{(a)}}{dx}\Big|_{x=t/2-t_1} = \frac{dh_z^{(1)}}{dx}\Big|_{x=t/2-t_1},$$
$$\frac{dh_y^{(a)}}{dx}\Big|_{x=-t/2+t_2} = \frac{dh_y^{(2)}}{dx}\Big|_{x=-t/2+t_2}, \quad \frac{dh_z^{(a)}}{dx}\Big|_{x=-t/2+t_2} = \frac{dh_z^{(2)}}{dx}\Big|_{x=-t/2+t_2}.$$
(7)

The distribution of the magnetic field describing by the set of Eqs. (1) and (4) can be written in the form

$$h_y^{(a)} = \cos\theta[A_1\sinh(k_0 x) + A_2\cosh(k_0 x)] + \sin\theta[A_3\sinh(k_1 x) + A_4\cosh(k_1 x)],$$

$$h_z^{(a)} = \sin\theta[A_1\sinh(k_0 x) + A_2\cosh(k_0 x)] - \cos\theta[A_3\sinh(k_1 x) + A_4\cosh(k_1 x)], \quad (8)$$

$$k_0 = (1-i)/\delta_0, \quad k_1 = (1-i)(\mu_1 + 1)^{1/2}/\delta_0,$$

$$h_y^{(j)} = \cos\varphi[B_1^{(j)}\sinh(k_0 x) + B_2^{(j)}\cosh(k_0 x)] + \sin\varphi[B_3^{(j)}\sinh(k_2 x) + B_4^{(j)}\cosh(k_2 x)],$$

$$h_z^{(j)} = \sin\varphi[B_1^{(j)}\sinh(k_0 x) + B_2^{(j)}\cosh(k_0 x)] - \cos\varphi[B_3^{(j)}\sinh(k_2 x) + B_4^{(j)}\cosh(k_2 x)], \quad (9)$$

$$k_2 = (1-i)(\mu_2 + 1)^{1/2}/\delta_0.$$

General solution (8) and (9) contains 12 constants, which can be found from Eqs. (6) and (7). After that, the amplitude $V_c$ of the pickup coil voltage can be calculated as[16,21]

$$V_c = Z_c I_0 = (2\pi/c)NI_0\left(\zeta_{yz}\big|_{x=t/2} + \zeta_{yz}\big|_{x=-t/2}\right), \quad (10)$$

where $N$ is the number of turns in the pickup coil and $\zeta_{yz}$ is the off-diagonal component of the surface impedance tensor, which can be expressed as[13,16]

$$\zeta_{yz}\big|_{\pm x=t/2} = -\frac{c}{4\pi\sigma} \times \frac{dh_z^{(j)}/dx}{h_y^{(j)}}\bigg|_{x=\pm t/2}. \quad (11)$$

The calculated field dependence of the off-diagonal magnetoimpedance $Z_c$ is shown in Fig. 4(a) at $f=500$ kHz for different values of the bias field $H_b$. The values of $Z_c$ are reduced to the dc ribbon resistance $R_{dc}=l/\sigma wt$, where $l$ is the ribbon length. In the model proposed, we assume that the ribbon has a single-domain structure. However, it is well known that at low external fields, a domain structure in the ribbon may exist. In particular, the observed hysteresis of the field dependence of the impedance denotes the appearance of the domain structure. To calculate the off-diagonal magnetoimpedance, we assume that the stripe domain structure exists at low external fields and we average the off-diagonal impedance response over the domain structure. The details of the averaging procedure can be found in Ref. 23. It should be noted that the presence of the domain structure reduces the off-diagonal impedance due to the contribution with opposite sings from the domains with different transverse magnetization direction. For example, in Co-based amorphous wires with the regular bamboo domain structure the voltage response averaged over domains tends to zero, and the off-diagonal impedance observed is very small and irregular.[17] In this case, dc bias current is needed to eliminate the domain structure and to observe the field-dependent signal in the pickup coil. In the studied samples, the anisotropy axis in the amorphous bulk deviates from the transverse direction in the ribbons. Moreover, the bias field changes the relative volume of two types of domains.[23] These two factors result in nonzero off-diagonal magnetoimpedance even in the presence of domain structure, since the signals from the domains are not compensated completely. Note also that we assume in calculations that the saturation magnetization, conductivity, and damping parameter are the same for the amorphous and crystalline regions. Corresponding modifications taking into account the variations of these parameters can be readily made in the framework of the present approach.

It follows from Fig. 4(a) that the field dependence of the off-diagonal impedance $Z_c$ shows the asymmetric two-peak behavior. The asymmetry growths with the bias field $H_b$, the negative field peak decreases and the positive field peak increases. The peak values of $H_e$ shift to the direction of the annealing field with the increase of the effective bias field. It is seen from Fig. 4(b) that the asymmetry in the field dependence of the off-diagonal impedance increases with the deviation of the unidirectional anisotropy field from the annealing field direction.

Figure 5 shows the effect of the uniaxial anisotropy angle $\psi$ on the field dependence of the off-diagonal impedance $Z_c$. The off-diagonal impedance and asymmetry between the peaks increase with the decrease of the anisotropy axis angle. However, the asymmetry disappears in the case of the transverse anisotropy, $\psi=0$. In this case, the field dependence of $Z_c$ becomes almost symmetric and just shifted with respect to the external field on the value of $H_b \sin\varphi$ (see Fig. 5). This result is similar to that obtained in the study of the GMI in amorphous wires with the circular anisotropy, when the presence of the dc bias current does not result in the asymmetry in the field dependence of the impedance.[16] Note that in the case of the transverse anisotropy in the amorphous bulk, the off-diagonal impedance differs from zero even in the presence of stripe domain structure, since the bias field changes the equilibrium magnetization angles in the domains and leads to difference in the relative volume of the domains.

Figure 6 illustrates the influence of the difference in the thickness of the surface crystalline layers on the off-diagonal magnetoimpedance. It can be easily shown that in the case of the same thickness of the surface layers, $t_1=t_2$, the magnetization distribution is symmetrical with respect to the ribbon center, and the off-diagonal impedance is equal to zero. It follows from Fig. 6 that the off-diagonal impedance $Z_c$ increases with the difference in the surface layers thickness $t_1 - t_2$, and the field dependence of $Z_c$ remains the same.

The comparison of the calculated field dependence of the off-diagonal impedance with the experimental data is shown in Fig. 2. It follows from Fig. 2 that the calculated dependences are in a qualitative agreement with the experimental results, and the calculated and measured values of the off-diagonal impedance are of the same order of magnitude. However, some disagreements are clearly seen from Fig. 2. The asymmetry between peaks in the experimental dependences at $f = 100$ kHz and $f = 500$ kHz is more pronounced as compared with the calculated results. Furthermore, within the region of the negative fields, the measured off-diagonal impedance drops more sharply with the field decrease. These disagreements may be related to the approximations made in the model. In particular, we neglect the coordinate dependence of the bias field, which allows one to find the solution for the off-diagonal impedance. Taking into account the spatial distribution of the bias field may be essential for a detail description of the asymmetric off-diagonal impedance field dependence. Moreover, to explain the observed hysteresis of the off-diagonal magnetoimpedance, an analysis of the effect of the domain structure should be done. Nevertheless, the approach developed allows one to describe the main features of the experimental data.

It should be noted that for the studied field-annealed amorphous ribbons, the impedance response measured as the voltage across the sample changes drastically with the frequency. At low frequencies, the single-peak "GMI valve" has been observed, whereas at high frequencies the GMI profile shows asymmetric two-peak behavior.[9] The transition from the single-peak to two-peak field dependence of the impedance can be explained as follows.[23] At low frequencies, the main contribution to the permeability is due to the domain-walls motion, and in this case the GMI response has the single-peak behavior. At high frequencies, the domain-walls motion is damped by eddy currents, and the magnetization rotational process determines the permeability, which results in the two-peak GMI profile.

On the contrary, the field dependence of the off-diagonal magnetoimpedance shows the two-peak behavior even at low frequencies, since the domain-wall motion contribution to the off-diagonal impedance is sufficiently small. This fact can be understood in the following way. The contribution from the domain-walls motion to the transverse permeability, $\mu_{tr}$, which is responsible for the GMI measured as the voltage across the sample, is given by[23–25]

$$\mu_{tr} = 1 + 4\pi\chi_0(\cos\theta_1 - \cos\theta_2)^2/(1 - i\omega/\omega_{dw}), \tag{12}$$

where $\chi_0$ is the static domain-walls susceptibility, $\omega_{dw}$ is the relaxation frequency for the domain-walls motion, and $\theta_1$ and $\theta_2$ are the equilibrium magnetization angles in the domains, which can be found from Eq. (3). Following the procedure described in Ref. 23, we can present the contribution from the domain-walls motion to the nondiagonal component of the permeability, $\mu_{od}$, in the form

$$\mu_{od} = 4\pi\chi_0(\cos\theta_1 - \cos\theta_2)(\sin\theta_1 - \sin\theta_2)/(1 - i\omega/\omega_{dw}). \tag{13}$$

It follows from the comparison of Eqs. (12) and (13) that $\mu_{tr}$ and $\mu_{od}$ have different dependence on the external magnetic field. Note that in the case of the transverse anisotropy without the bias field the nondiagonal permeability component, $\mu_{od}$, equals zero.[24] The calculations by means of Eqs. (3), (12), and (13) show that for typical parameters of the studied ribbons the nondiagonal component $\mu_{od}$ is less than $\mu_{tr}$ by one order of the magnitude. Moreover, the antisymmetrical transverse magnetic field distribution over the ribbon thickness leads to a contribution with opposite signs to the pickup coil voltage from two parts of the sample, $x>0$ and $x<0$. Hence, the contribution from the domain-walls motion to the off-diagonal impedance is proportional to the difference in the surface layers thickness, $t_1-t_2$, only, whereas the GMI measured as the voltage drop across the sample is proportional to the ribbon thickness. Thus, we can conclude that the contribution from the domain-walls motion to the off-diagonal magnetoimpedance can be neglected even at low frequencies.

In concluding of this section, we assume that the asymmetry in the magnetization distribution over the ribbon thickness is related to the different thickness of the surface crystalline layers, which may be attributed to peculiarities of the field-annealing process. The model allows to one to describe the field dependence of the off-diagonal impedance by using reasonable values of the fitting parameters. Note that the asymmetry may occur also due to other reasons, in particular due to the variation of the anisotropy field and the anisotropy axis angle over the ribbon thickness. Further investigation of the effect of the annealing conditions on the off-diagonal magnetoimpedance should be performed, and this study is now in progress.

## IV. CONCLUSIONS

The results of the study of the off-diagonal magnetoimpedance in field-annealed Co-based amorphous ribbons are presented. The field dependence of the off-diagonal impedance shows asymmetric two-peak behavior. The asymmetry is related to the formation of the hard magnetic phase due to the surface crystallization of the amorphous ribbons. To explain experimental results, the model based on the concept of the surface impedance tensor is developed. It is assumed that the ribbon consists of the amorphous region and two surface crystalline layers having different thickness. The coupling between the surface crystalline layers and the amorphous phase is described in terms of the effective bias field, which is assumed to be constant over the ribbon volume. The calculated dependences describe qualitatively the experimental data. The results obtained may be useful to develop sensitive magnetic-field sensors.

## ACKNOWLEDGMENTS

The authors would like to thank Dr. L. Kraus for fruitful discussions. This work was supported by the Korea Science and Engineering Foundation through ReCAMM. N.A.B. acknowledges the support of the Brain Pool Program.

**References**

[1] L. V. Panina, J. Magn. Magn. Mater. **249**, 278 (2002).

[2] L. Kraus, Sensors Actuators A **106**, 187 (2003).

[3] T. Kitoh, K. Mohri, and T. Uchiyama, IEEE Trans. Magn. **31**, 3137 (1995).

[4] S. H. Song, K. S. Kim, S. C. Yu, C. G. Kim, and M. Vazquez, J. Magn. Magn. Mater. **215–216**, 532 (2000).

[5] D. P. Makhnovskiy, L. V. Panina, and D. J. Mapps, Appl. Phys. Lett. **77**, 121 (2000).

[6] C. G. Kim, K. J. Jang, H. C. Kim, and S. S. Yoon, J. Appl. Phys. **85**, 5447 (1999).

[7] K. J. Jang, C. G. Kim, S. S. Yoon, and K. H. Shin, IEEE Trans. Magn. **35**, 3889 (1999).

[8] Y. W. Rheem, C. G. Kim, C. O. Kim, and S. S. Yoon, J. Appl. Phys. **91**, 7433 (2002).

[9] C. G. Kim, C. O. Kim, and S. S. Yoon, J. Magn. Magn. Mater. **249**, 293 (2002).

[10] K. H. Shin, C. D. Graham Jr., and P. Y. Zhou, IEEE Trans. Magn. **28**, 2772 (1992).

[11] C. G. Kim, Y. W. Rheem, C. O. Kim, E. E. Shalyguina, and E. A. Ganshina, J. Magn. Magn. Mater. **262**, 412 (2003).

[12] A. Antonov, I. Iakubov, and A. Lagarkov, IEEE Trans. Magn. **33**, 3367 (1997).

[13] N. A. Usov, A. S. Antonov, and A. N. Lagar'kov, J. Magn. Magn. Mater. **185**, 159 (1998).

[14] A. S. Antonov, I. T. Iakubov, and A. N. Lagarkov, J. Magn. Magn. Mater. **187**, 252 (1998).

[15] L. V. Panina, K. Mohri, and D. P. Makhnovskiy, J. Appl. Phys. **85**, 5444 (1999).

[16] D. P. Makhnovskiy, L. V. Panina, and D. J. Mapps, Phys. Rev. B **63**, 144424 (2001).

[17] S. Sandacci, D. Makhnovskiy, L. Panina, K. Mohri, and Y. Honkura, IEEE Trans. Magn. **40**, 3505 (2004).

[18] A. S. Antonov and I. T. Iakubov, J. Phys. D **32**, 1204 (1999).

[19] N. Fry, D. P. Makhnovskiy, L. V. Panina, S. T. Sandacci, D. J. Mapps, and M. Akhter, IEEE Trans. Magn. **40**, 3358 (2004).

[20] D. P. Makhnovskiy, A. S. Antonov, A. N. Lagar'kov, and L. V. Panina, J. Appl. Phys. **84**, 5698 (1998).

[21] L. V. Panina, D. P. Makhnovskiy, and K. Mohri, J. Magn. Magn. Mater. **272–276**, 1452 (2004).

[22] N. A. Buznikov, C. G. Kim, C. O. Kim, and S. S. Yoon, J. Magn. Magn. Mater. **288**, 130 (2005).

[23] N. A. Buznikov, C. G. Kim, C. O. Kim, and S. S. Yoon, Appl. Phys. Lett. **85**, 3507 (2004).

[24] L. V. Panina, K. Mohri, T. Uchiyama, M. Noda, and K. Bushida, IEEE Trans. Magn. **31**, 1249 (1995).

[25] F. L. A. Machado and S. M. Rezende, J. Appl. Phys. **79**, 6558 (1996).

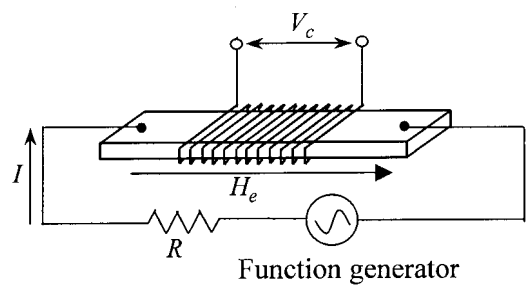

FIG. 1. Schematic diagram of the experimental setup.

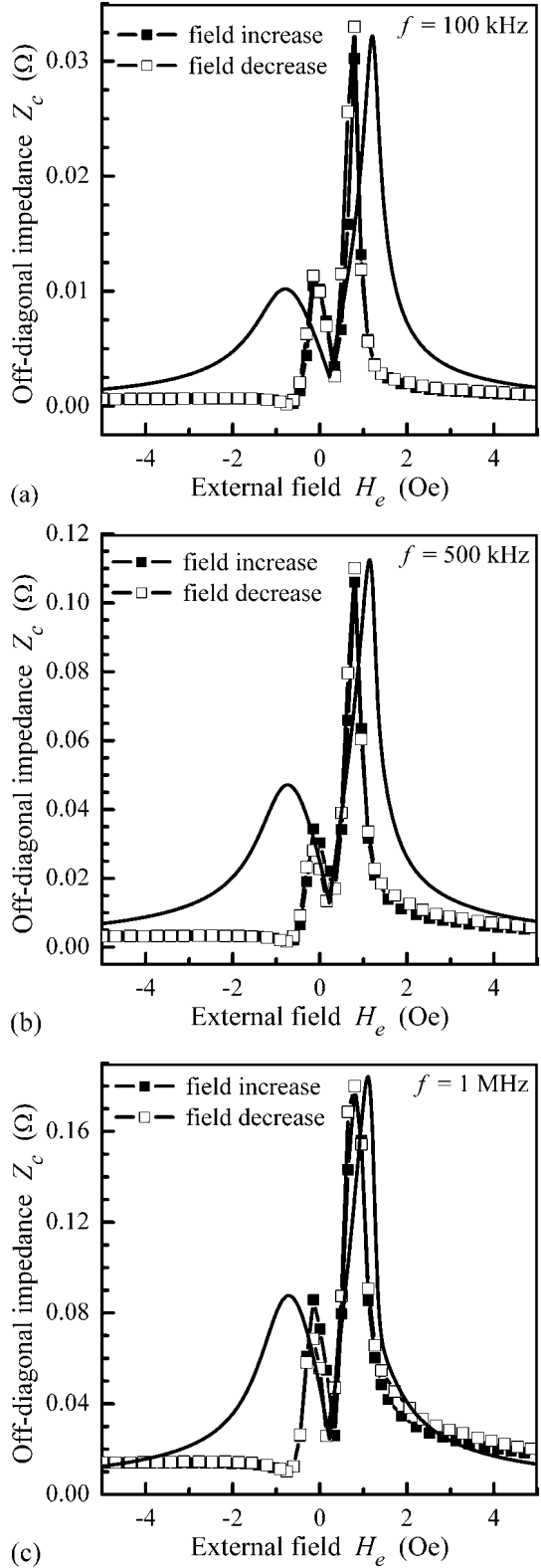

FIG. 2. Dependence of the off-diagonal impedance $Z_c$ on the external field $H_e$ at current frequency $f = 100$ kHz (a), $f = 500$ kHz (b), and $f = 1$ MHz (c). Symbols, experimental data; solid line, calculations. Parameters used for calculations are $M = 600$ G, $H_a = 1$ Oe, $H_u = 200$ Oe, $H_b = 0.3$ Oe, $\psi = 0.05\pi$, $\varphi = 0.35\pi$, $\sigma = 10^{16}$ s$^{-1}$, $t_1 = 1$ μm, $t_2 = 0.65$ μm, and $\kappa = 0.1$.

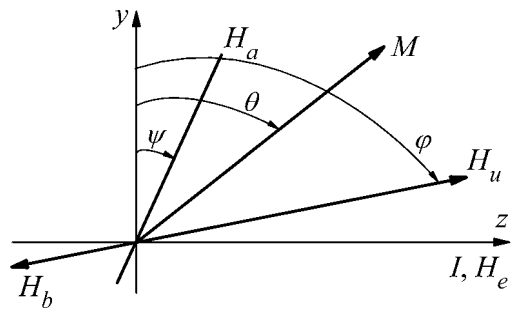

FIG. 3. A sketch of the coordinate system used for analysis. All the vectors lie within the $y$–$z$ plane.

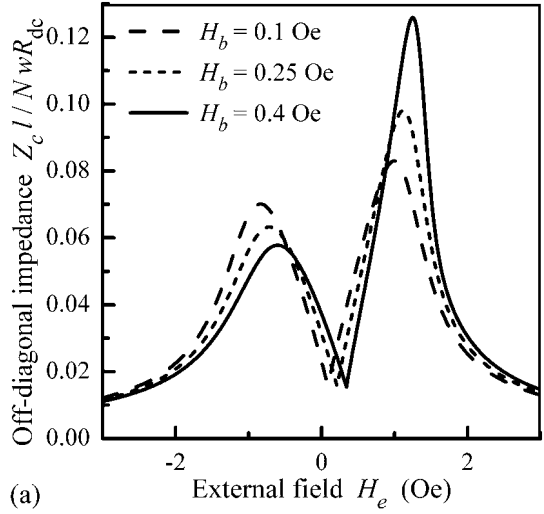

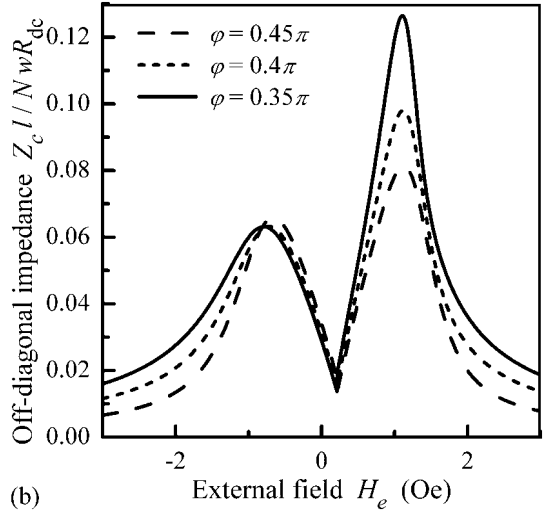

FIG. 4. Off-diagonal impedance $Z_c$ vs external field $H_e$ at $\varphi=0.4\pi$ and different $H_b$ (a) and $H_b=0.25$ Oe and different $\varphi$ (b). Parameters used for calculations are $M=600$ G, $H_a=1$ Oe, $H_u=200$ Oe, $f=500$ kHz, $\psi=0.05\pi$, $\sigma=10^{16}$ s$^{-1}$, $t=20$ μm, $t_1=1$ μm, $t_2=0.5$ μm, and $\kappa=0.1$.

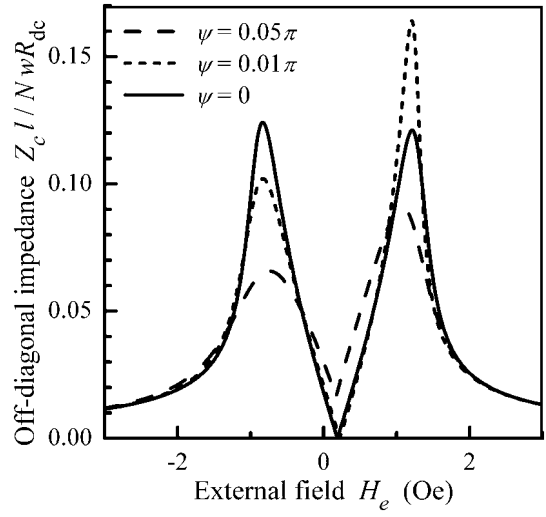

FIG. 5. Off-diagonal impedance $Z_c$ vs external field $H_e$ at different $\psi$. Parameters used for calculations are $M=600$ G, $H_a=1$ Oe, $H_u=200$ Oe, $H_b=0.2$ Oe, $f=500$ kHz, $\varphi=0.4\pi$, $\sigma=10^{16}$ s$^{-1}$, $t=20$ μm, $t_1=1$ μm, $t_2=0.5$ μm, and $\kappa=0.1$.

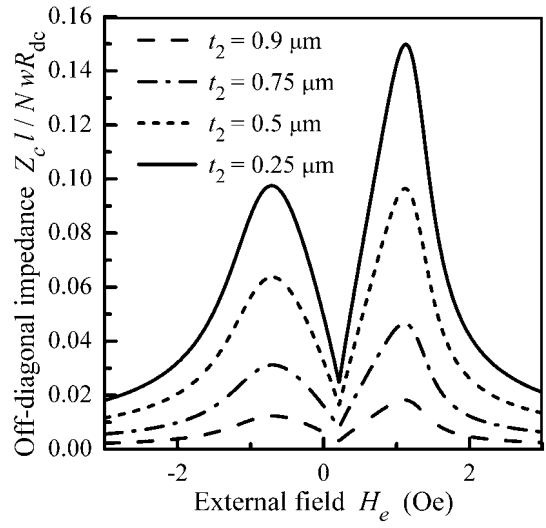

FIG. 6. Off-diagonal impedance $Z_c$ vs external field $H_e$ at $t_1=1$ μm and different $t_2$. Parameters used for calculations are $M=600$ G, $H_a=1$ Oe, $H_u=200$ Oe, $H_b=0.25$ Oe, $f=500$ kHz, $\psi=0.05\pi$, $\varphi=0.4\pi$, $\sigma=10^{16}$ s$^{-1}$, $t=20$ μm, and $\kappa=0.1$.